\documentclass{aastex631}
\usepackage{subfigure}
\usepackage{makecell}
\usepackage{graphicx}	
\usepackage{multirow}
\usepackage{epstopdf}
\usepackage[normalem]{ulem}
\usepackage[figuresright]{rotating}
\usepackage{array}
\usepackage{tabularx}
\useunder{\uline}{\ul}{}

\makeatletter

\newcommand{\Rmnum}[1]{\expandafter\@slowromancap\romannumeral #1@}

\newcommand{\kms}{km\,s$^{-1}$}
\newcommand{\degree}{$^\circ$}
\makeatother
\shorttitle{Sequential plasma ejections in tangled loops}
\shortauthors{Zuo et al.}
\graphicspath{{./}{}}

\begin{document}

\title{Sequential ejections of plasma blobs due to unbraiding of tangled loops in the solar atmosphere}

\correspondingauthor{Zhenghua Huang}
\email{z.huang@sdu.edu.cn}

\author{Xiuhui Zuo}
\affiliation{Shandong Key Laboratory of Optical Astronomy and Solar-Terrestrial Environment, Institute of Space Sciences, Shandong University, Weihai 264209, Shandong, China}
\affiliation{State Key Laboratory of Solar Activity and Space Weather, Chinese Academy of Sciences, Beijing, 100190, China}

\author[0000-0002-2358-5377]{Zhenghua Huang}
\affiliation{Shandong Key Laboratory of Optical Astronomy and Solar-Terrestrial Environment, Institute of Space Sciences, Shandong University, Weihai 264209, Shandong, China}
\affiliation{State Key Laboratory of Solar Activity and Space Weather, Chinese Academy of Sciences, Beijing, 100190, China}

\author[0000-0002-0210-6365]{Hengyuan Wei}
\affiliation{Shandong Key Laboratory of Optical Astronomy and Solar-Terrestrial Environment, Institute of Space Sciences, Shandong University, Weihai 264209, Shandong, China}

\author{Chao Zhang}
\affiliation{Shandong Key Laboratory of Optical Astronomy and Solar-Terrestrial Environment, Institute of Space Sciences, Shandong University, Weihai 264209, Shandong, China}

\author{Boyu Sun}
\affiliation{Shandong Key Laboratory of Optical Astronomy and Solar-Terrestrial Environment, Institute of Space Sciences, Shandong University, Weihai 264209, Shandong, China}

\author{Youqian Qi}
\affiliation{Shandong Key Laboratory of Optical Astronomy and Solar-Terrestrial Environment, Institute of Space Sciences, Shandong University, Weihai 264209, Shandong, China}

\author[0000-0002-8827-9311]{Hui Fu}
\affiliation{Shandong Key Laboratory of Optical Astronomy and Solar-Terrestrial Environment, Institute of Space Sciences, Shandong University, Weihai 264209, Shandong, China}

\author{Weixin Liu}
\affiliation{Shandong Key Laboratory of Optical Astronomy and Solar-Terrestrial Environment, Institute of Space Sciences, Shandong University, Weihai 264209, Shandong, China}

\author{Mingzhe Sun}
\affiliation{Shandong Key Laboratory of Optical Astronomy and Solar-Terrestrial Environment, Institute of Space Sciences, Shandong University, Weihai 264209, Shandong, China}

\author[0000-0001-9427-7366]{Ming Xiong}
\affiliation{State Key Laboratory of Solar Activity and Space Weather, Chinese Academy of Sciences, Beijing, 100190, China}
\affiliation{College of Earth and Planetary Sciences, University of Chinese Academy of Sciences, Beijing, 100049, China}

\author[0000-0001-8938-1038]{Lidong Xia}
\affiliation{Shandong Key Laboratory of Optical Astronomy and Solar-Terrestrial Environment, Institute of Space Sciences, Shandong University, Weihai 264209, Shandong, China}



\begin{abstract}
Nanoflares, which are consequences of braids in tangled magnetic fields, are an important candidate to heat the solar corona to million degrees.
However, their observational evidence is sparse and many of their observational characteristics are yet to be discovered.
With the high-resolution observations taken by the Extreme Ultraviolet Imager onboard the Solar Orbiter, here we study a series of ejections of plasma blobs resulted from a braided magnetic loops in the upper transition region and reveal some critical characteristics of such processes.
The cores of these ejections have a size of about 700\,km, a duration less than 1 minute and a speed of about 90\,\kms.
An important characteristic is that these plasma blobs are apparently constrained by the post-reconnection magnetic loops, along which they show an extension of up to about 2\,000\,km.
The propagation of unbraiding nodes along the main axis of the tangled loops has a speed of about 45\,\kms.
The separation angles between the post-reconnection loops and the main axis of the tangled loops are about 30\degree.
The observations from the Atmospheric Imaging Assembly reveal that the braiding loops are upper transition region structures.
Based on these observations, the typical magnetic free energy producing a blob is estimated to be about $3.4\times10^{23}$\,erg, well in the nano-flare regime, while the kinematic energy of a blob is about $2.3\times10^{23}$\,erg, suggesting that a majority of magnetic free energy in a magnetic braid is likely transferred into kinematic energy.
\end{abstract}

\section{Introduction}\label{sec:intro}
Back to the 1970s, Eugene Parker pointed out that the magnetic-field lines in the solar corona can be tangled due to the random motions at their footpoints, 
and this process can result in energy release in a nano-flare scale to heat the corona due to small-scale current sheets generated in the magnetic system with tangled field lines\,\citep{1972ApJ...174..499P,1983ApJ...264..635P,1983ApJ...264..642P,1988ApJ...330..474P}.
Such a scenario is so-call ``nano-flare'' or ``magnetic braiding'' model and it is believed to be one of the main mechanisms to heat the million-degree corona\,\citep[e.g.][and references therein]{2004ApJ...617L..85P,2006SoPh..234...41K,2009ApJ...696.1339W}.

\par
The characteristic scale of the energy release due to magnetic braidings in the solar corona is believed to be less than $10^3$\,km\,\citep{1988ApJ...330..474P}, and it decreases exponentially with increasing braiding complexity\,\citep{Wilmot_Smith_2009}.
Therefore, evidence of magnetic braids in the solar corona is not yet convincing due to the lack of observations with enough spatiotemporal resolutions\,\citep{2011ApJ...736....3V,2017ApJ...837..108P}.

\par
Using the observations with a spatial resolution of about 150 km and a cadence of 5 s from the High Resolution Coronal Imager\,\citep[Hi-C,][]{2014SoPh..289.4393K}, 
\citet{2013Natur.493..501C} reported observational evidence of spatially-resolved magnetic braids in the solar corona, showing that the spatial scales of the magnetic braids are approximately 150\,km.
In Hi-C observations, energy releases in magnetic braids are shown as a sequence of isolated bright dots along the tangled loops together with occurrence of plasma heating and accelerations along the loops.

\par
Later, the Interface Region Imaging Spectrograph\,\citep[IRIS,][]{2014SoPh..289.2733D} provides both imaging and spectral observations of the solar transition region with a spatial resolution of about 250\,km.
Combining IRIS data and those from the Solar Dynamics Observatory\,\citep[SDO,][]{2012SoPh..275....3P}, from imaging, spectral and magnetic aspects, \citet{2018ApJ...854...80H} reported a more comprehensive evidence of magnetic braids associated with jet activities in a coronal spiral structure.
Moreover, studies based on IRIS data have discovered perpendicular jets in some magnetic features, and these phenomena support the magnetic braiding scenario.
\citet{2020ApJ...899...19C} discovered mini-jets with lengths of 1--6 Mm, widths of 0.2--1 Mm, durations of 10--50 s and speeds of 100--350\,\kms\ occurred in an activated solar prominence.
Those mini-jets have directions perpendicular to the main extending axis of the prominence threads, suggesting that they were induced by internal magnetic reconnection in tangled prominence threads.
Similar jet events but named ``nano-jets'' were also found by \citet{2021NatAs...5...54A}, who also perform a numerical simulation to confirm that these jets are products of slingshot effect from reconnections of tangled and curved magnetic field lines.

\par
Since the launch of the Solar Orbiter\,\citep[SO,][]{2020A&A...642A...1M} in 2020, images of the solar corona with spatial resolutions up to 100\,km/pixel and cadences of a few seconds become available.
Based on the data retrieved by the Extreme Ultraviolet Imager\,\citep[EUI,][]{2020A&A...642A...8R} at 174\,\AA\ passband and spatial resolutions of 200--250 km and cadences of 3--10\,s, \citet{2022A&A...667A.166C} reported observations of untangling of small-scale coronal braids and their consequences of impulsive coronal heating events in four active regions.
They also inferred that coronal magnetic braids are more common in structures overlying cooler chromospheric filamentary structures.
\citet{2022A&A...667A.166C} also claimed some signatures of ``nano-jets'' in those braiding magnetic systems, though they are not as clear as those in the IRIS observations.

\par
In the present study, we exploit the high resolution data from EUI/SO to investigate dynamics resulted from tangled magnetic flux in the solar corona.
We observe plasma ejections due to relaxations of tangled coronal loops.
These ejections are showing as bright blobs propagating in the direction in perpendicular to the loop lengths, suggesting that the magnetic tensions are at work.
Interestingly, the ejecta also show clear extensions along the loops, evidencing that the heated plasma is still constrained by the magnetic field.
These observations not only show another observational evidence of magnetic braids in the solar atmosphere, but also clearly point out that direct energy dissipation in magnetic braids is likely constrained by the magnetic field, and therefore extra mechanisms are required to transfer energy from the unbraiding site to the ambient corona.

\par
The remain of the paper is organised as follows. 
The data are described in Section\,\ref{sec:obs}. 
The results are presented in Section\,\ref{sec:res}. 
The discussion is given in Section\,\ref{sec:dis}.
And the conclusions are given in Section\,\ref{sec:con}.

\section{Observations} \label{sec:obs}
\begin{figure*}
	\centering
	\includegraphics[width=\textwidth]{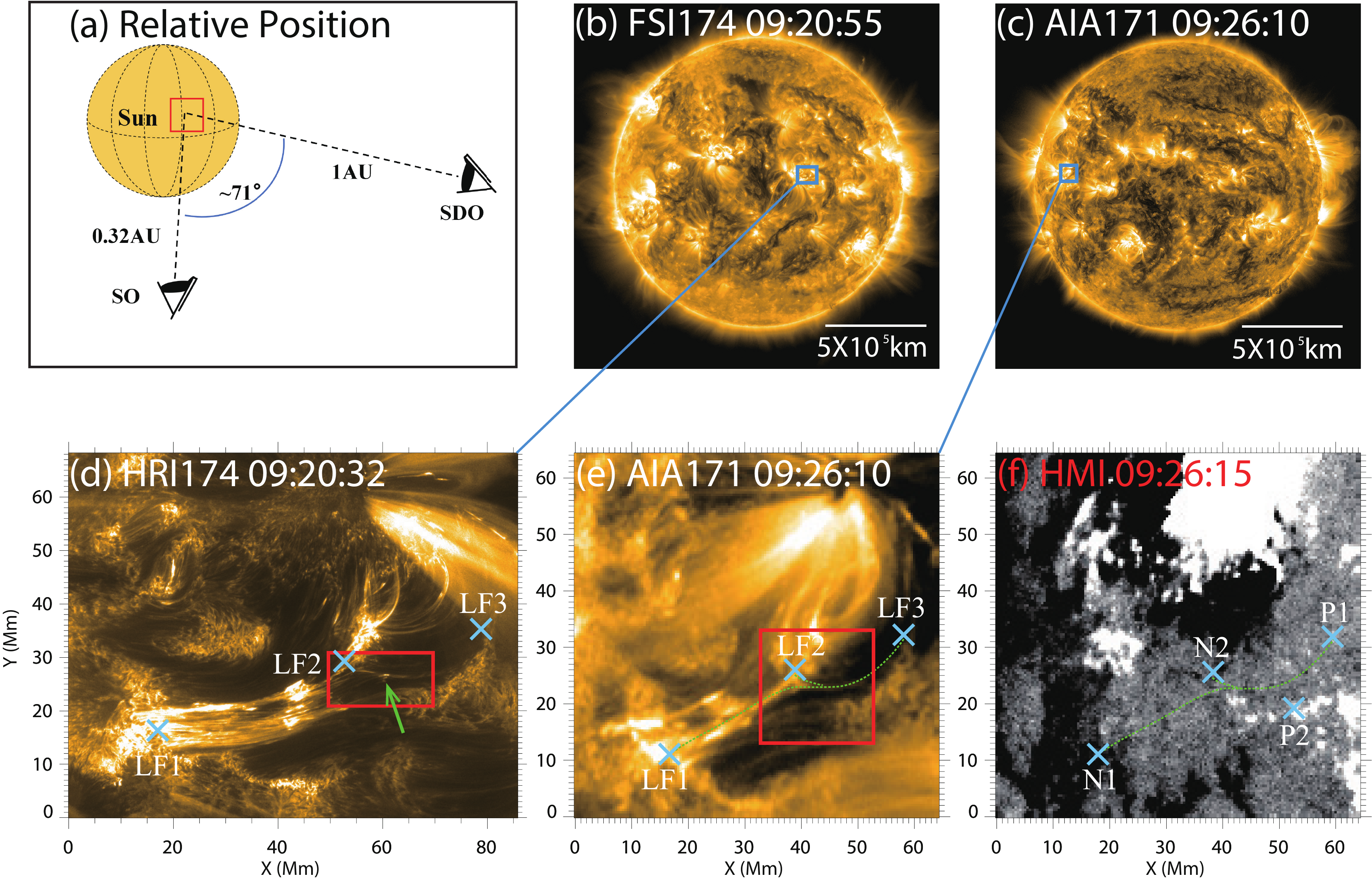}
	\caption{Overviews to the observations studied here.
	(a): Viewing perspectives of the sun by SO and SDO.
	(b): The full disk of the sun in EUI 174\,\AA\ passband, on which the square in blue indicates the region containing the region of interest.
	(c): The full disk of the sun in AIA 171\,\AA\ passband, on which the region of interest is located near the east limb of the disk (the square in blue).
	(d): The region of interest in high-resolution image of EUI 174\,\AA\ passband.
	On this map, the square in red indicates the field-of-view shown in Figure\,\ref{fig2} and the green arrow points to the location where the activities take place.
	(e): The region of interest in high-resolution image of AIA 171\,\AA\ passband, on which the square in red indicates the field-of-view shown in Figure\,\ref{fig4}.
	The connections of the loops in the system are outlined by the green dotted lines in panel (e) and their major footpoints are denoted by cyan cross symbols with remarks of ``LF1'', ``LF2'' and ``LF3''.
	(f): The HMI magnetogram of the region of interest scaled from --50\,G (black) to 50\,G (white).
	The connections of the loops outlined in panel (e) are overplotted and the polarities associated with the loop system are marked by cyan cross symbols with remarks of ``N1'' and ``N2'' (negative) and ``P1'' and ``P2'' (positive).
	\label{fig1}}
\end{figure*}

The data utilized in this study were taken on October 13, 2023, and the region-of-interest (ROI) is part of the active region of NOAA 13465.
The observations from EUI/SO, AIA/SDO and the Helioseismic and Magnetic Imager \citep[HMI,][]{2012SoPh..275..207S} onboard SDO are studied here.
In Figure\,\ref{fig1}, we show the overview of the observations.
The separation angle between SO and SDO is about 71\degree(Figure\,\ref{fig1}(a)).
While the data were taken, SO was 0.32\,AU away from the Sun and its solar latitude was 2.86\degree.
The sunlight travel time to SO is about 340\,s ahead that to SDO.
The ROI is located near the center of the solar disk as viewed by SO, while that is near the limb in the SDO views (Figure\,\ref{fig1}(b)\&(c)).
The coalignment between the SDO and SO data are firstly assessed based on the viewing angles of the two satellites and then optimized manually by cross-checking characteristic features in the region (such as footpoints and intersections of coronal loops).

\par
The observations from the High-Resolution Imager (HRI) of the EUI/SO was spanning from 09:10\,UT to 09:30\,UT.
The data from the HRI/EUI 174\,\AA\ passband are analysed, which has a pixel size of 0.492\arcsec\ (i.e. 110\,km/pixel) and a cadence of 6\,s.
An example of the HRI/EUI observations is given in Figure\,\ref{fig1}(d), which shows great details of the coronal features.
The AIA/SDO data from the EUV passbands of 94\,\AA, 131\,\AA, 171\,\AA, 193\,\AA, 211\,\AA, 304\,\AA\ and 335\,\AA\ are also analysed.
The AIA data taken at 1\,AU have a spatial resolution of 0.6\arcsec\ per pixel, which is equivalent to 426\,km per pixel.
The cadence of the AIA data is 12\,s.
A snapshot of the ROI viewed in AIA 171\,\AA\ is shown in Figure\,\ref{fig1}.
The AIA data show much less details of the features, but the data of multiple passbands can provide some diagnostics on the thermal distributions of the region.
An HMI/SDO longitudinal magnetogram of the region is shown in Figure\,\ref{fig1}(f), which has a spatial resolution of 0.6\arcsec\ per pixel.
Although the ROI locates near the limb in the HMI field-of-view, the magnetogram can still give some general information of the magnetic connection of the active features studied in the paper (see the remarks in Figure\,\ref{fig1}(e)\&(f)).


\begin{figure*}
	\centering
	\includegraphics[width=0.95\textwidth]{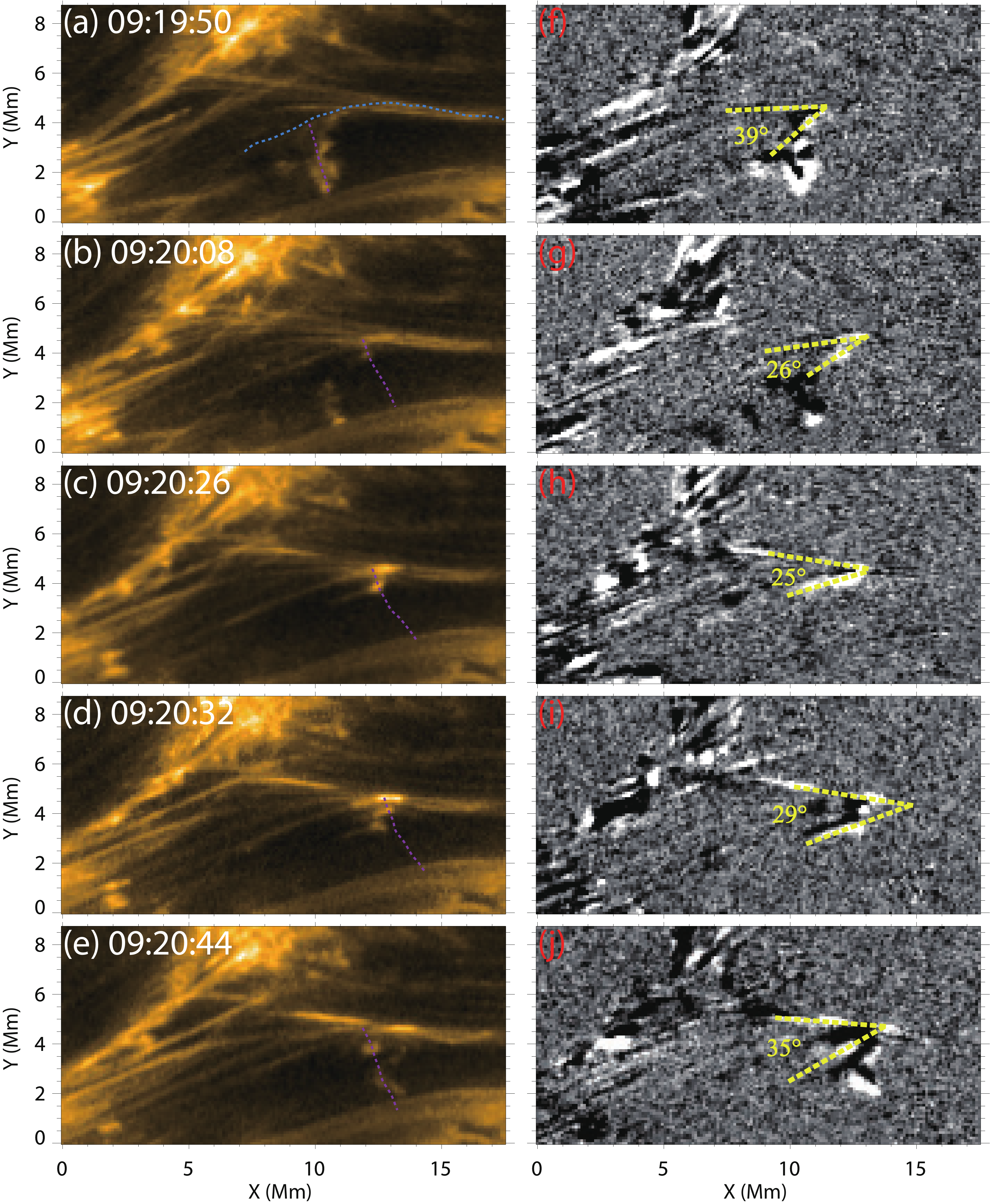}
	\caption{Still images of five ejections of the HRI/EUI 174\,\AA\ passband when they are best seen.
	The left column (panels (a)--(e)) shows the original images when the ejections are clearly seen, while the right column (panels (f)--(j)) shows the running difference images.
	The blue dashed line in panel (a) draws the path of the main axis of the loop system, along which the time-slice map in Figure\,\ref{fig3}(f) is obtained.
	The purple dashed lines in panels (a)--(e) draw the paths of the ejections, from which the time-slice maps in Figure\,\ref{fig3}(a)--(e) are obtained.
	The angles marked on the running different images show the angles between the main axis and the expelled loops.
	An animation showing the evolution of this region in original and differential images of HRI/EUI 174\,\AA\ is given online. 
	\label{fig2}}
\end{figure*}

\begin{figure*}
	\centering
	\includegraphics[width=\textwidth]{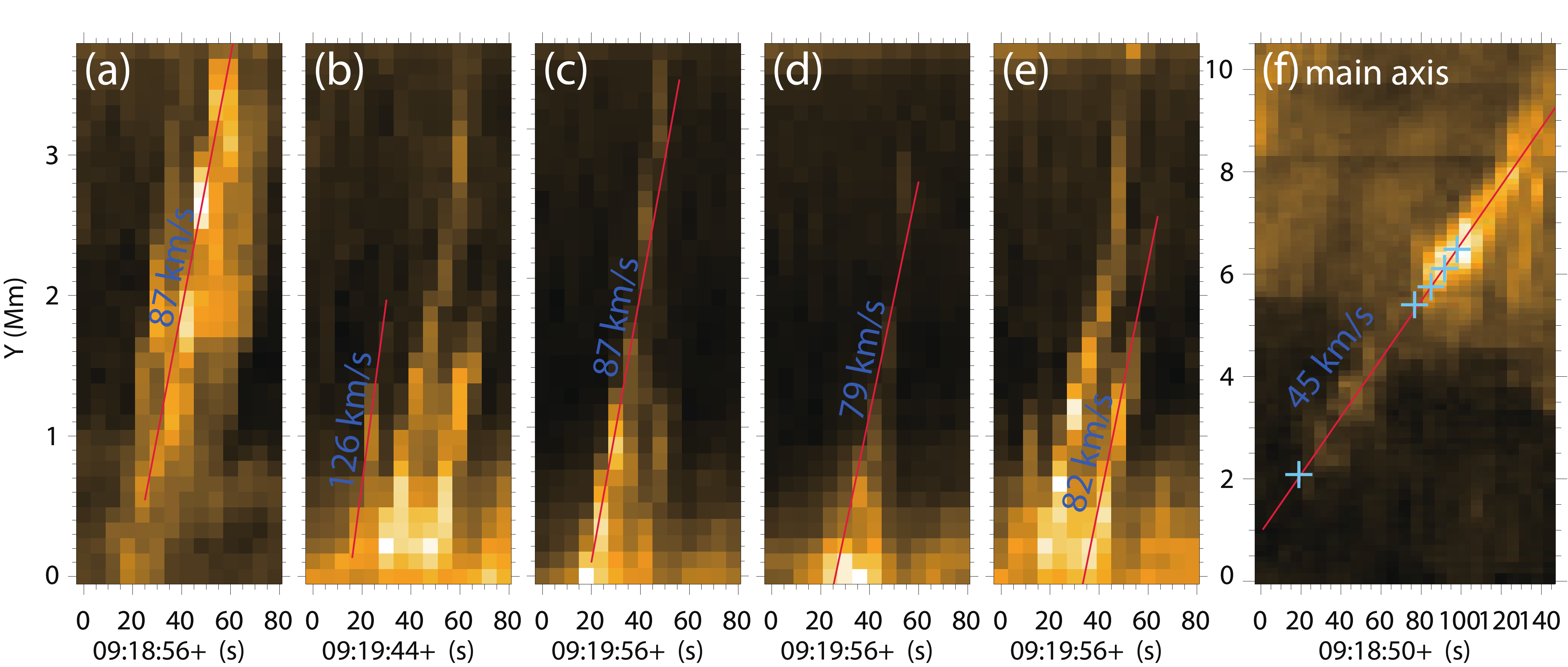}
	\caption{Time-slice (T-S) maps of the five ejections and the main axis of the loop system based on the HRI/EUI observations.
	From panel (a) to panel (e), the T-S maps of ejection-1 to ejection-5 are shown.
	The T-S map of the main axis is shown in panel (f).
	The speeds of the ejections of the plasma blobs and bright node along the main axis are marked, which are derived from the slopes of their tractories on these maps (the solid lines in red). 
	The cyan plus symbols in panel (f) mark the time and locations where the five ejections start.
	}
	\label{fig3}
\end{figure*}

\begin{figure*}
	\centering
	\includegraphics[width=\textwidth]{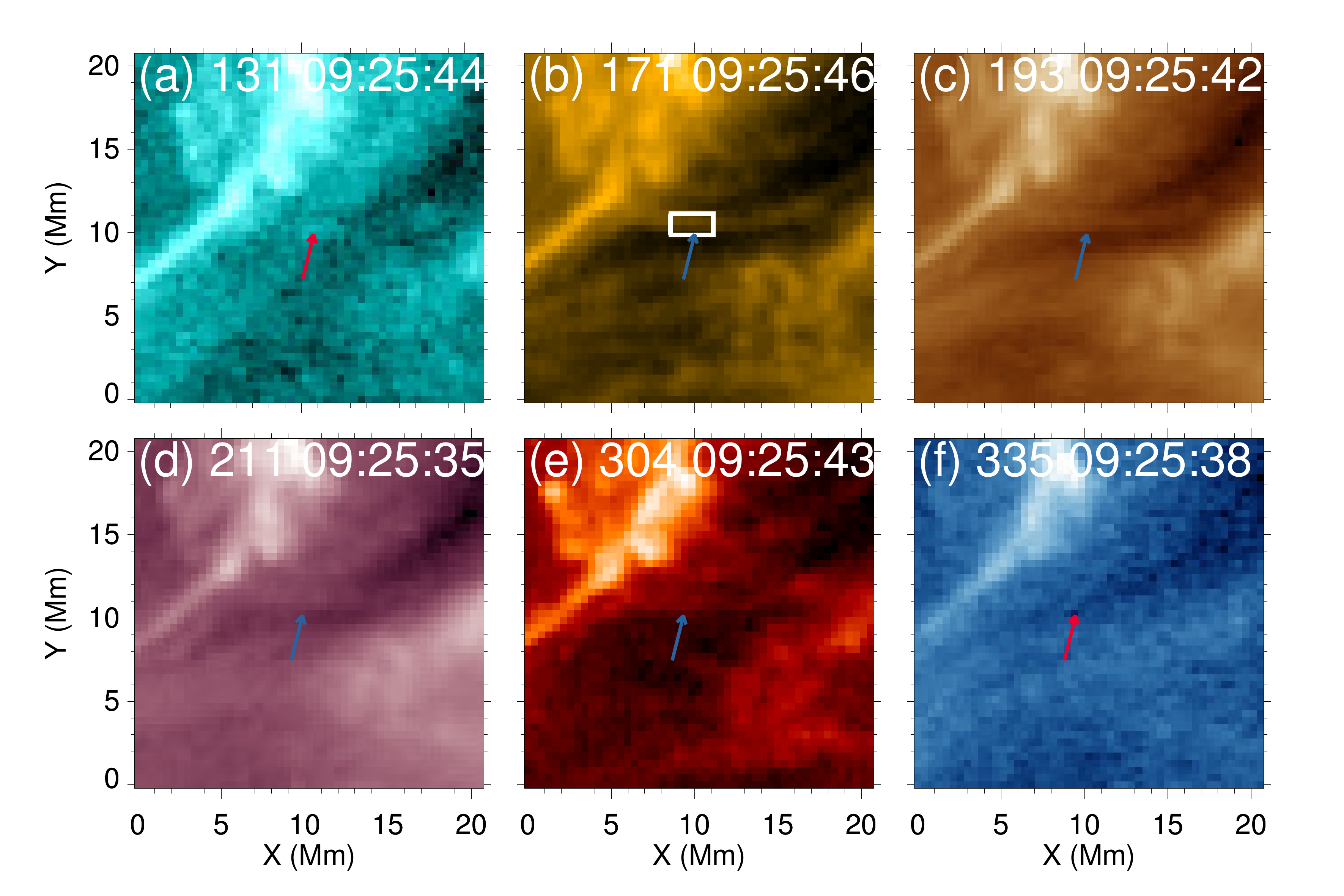}
	\caption{The region of interest viewed in the AIA 131\,\AA\ (a), 171\,\AA\ (b), 193\,\AA\ (c), 211\,\AA\ (d), 304\,\AA\ (e) and 335\,\AA\ (f) passbands.
	The arrows point to the location where the bright node in the main axis of the loop system is located.
	The white square denotes the region, from which the lightcurves in Figure\,\ref{fig5} are obtained.
	An animation showing the evolution of this region in these AIA  passbands is given online.
	}
	\label{fig4}
\end{figure*}

\begin{figure*}
	\centering
	\includegraphics[width=15cm]{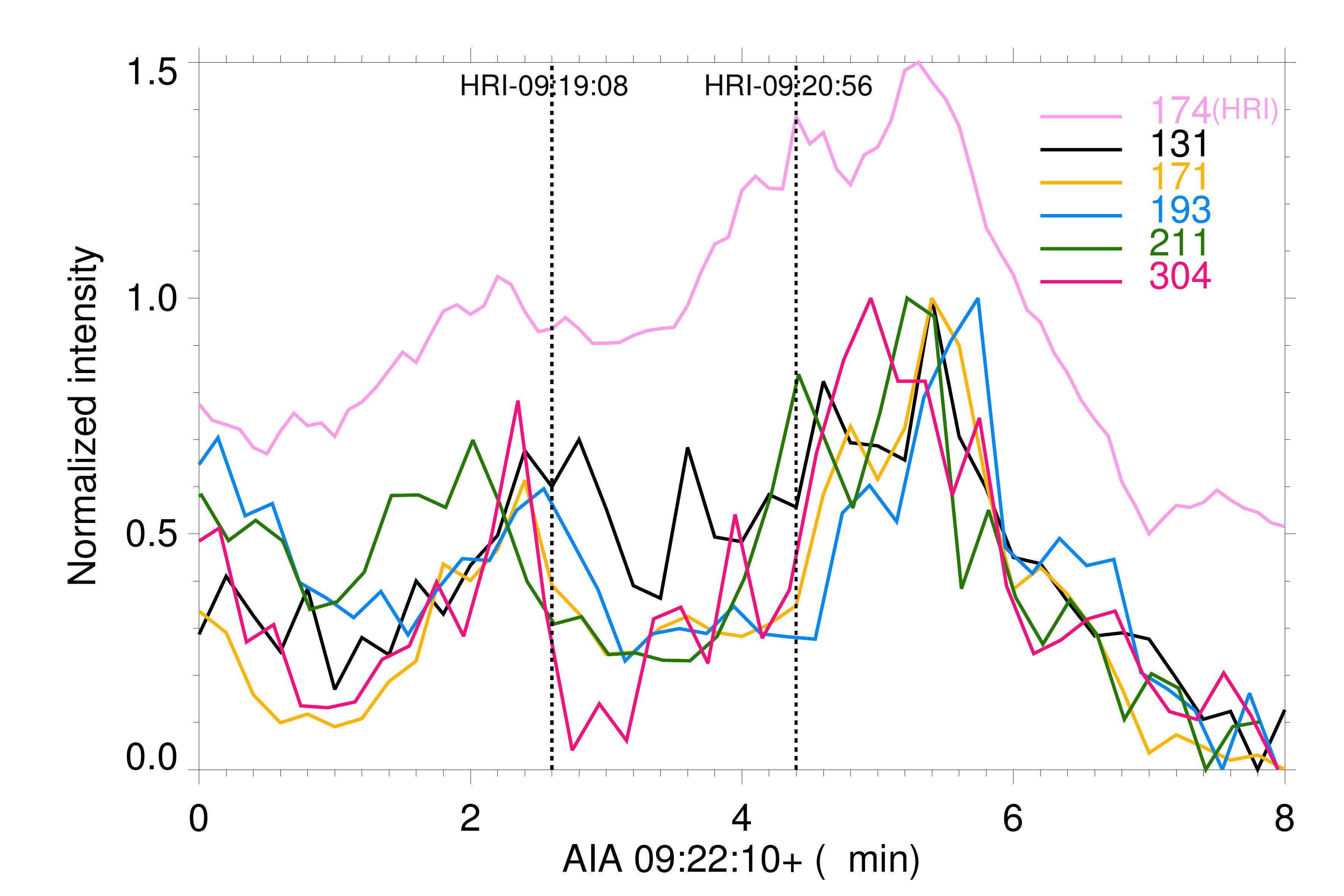}
	\caption{The lightcurves of the bright node in the main axis of the loop system in some AIA passbands, and that of HRI/EUI 174\,\AA\ summed from the main axis and shifted to the AIA time domain.
	The time range between the two dashed lines is the period when the five ejections take place.
	}
	\label{fig5}
\end{figure*}

\begin{figure*}
	\centering
	\includegraphics[width=\textwidth]{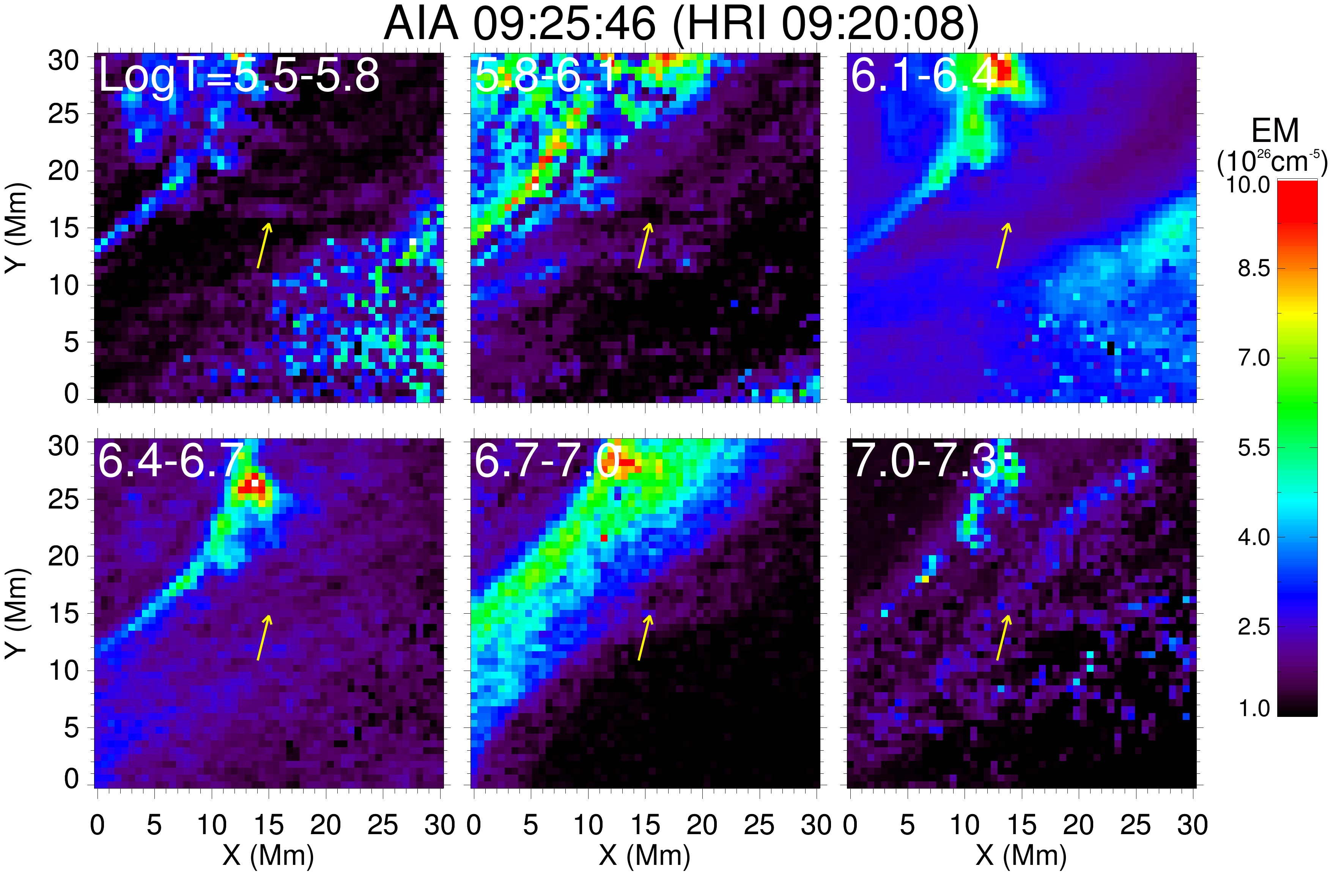}
	\caption{The EM maps of the region of interest derived from the AIA EUV images for the temperature ranges of logT/K=5.5--5.8, 5.8--6.1, 6.1--6.4, 6.4--6.7, 6.7--7.0 and 7.0--7.3.
	The arrows point to the locations of the bright node at the main axis, where we can see the structure only in the EM map of logT/K=5.5-5.8 and logT/K=6.1-6.4.
	An associated animation is given online.
	}
	\label{fig6}
\end{figure*}
\begin{figure*}
	\centering
	\includegraphics[width=\textwidth]{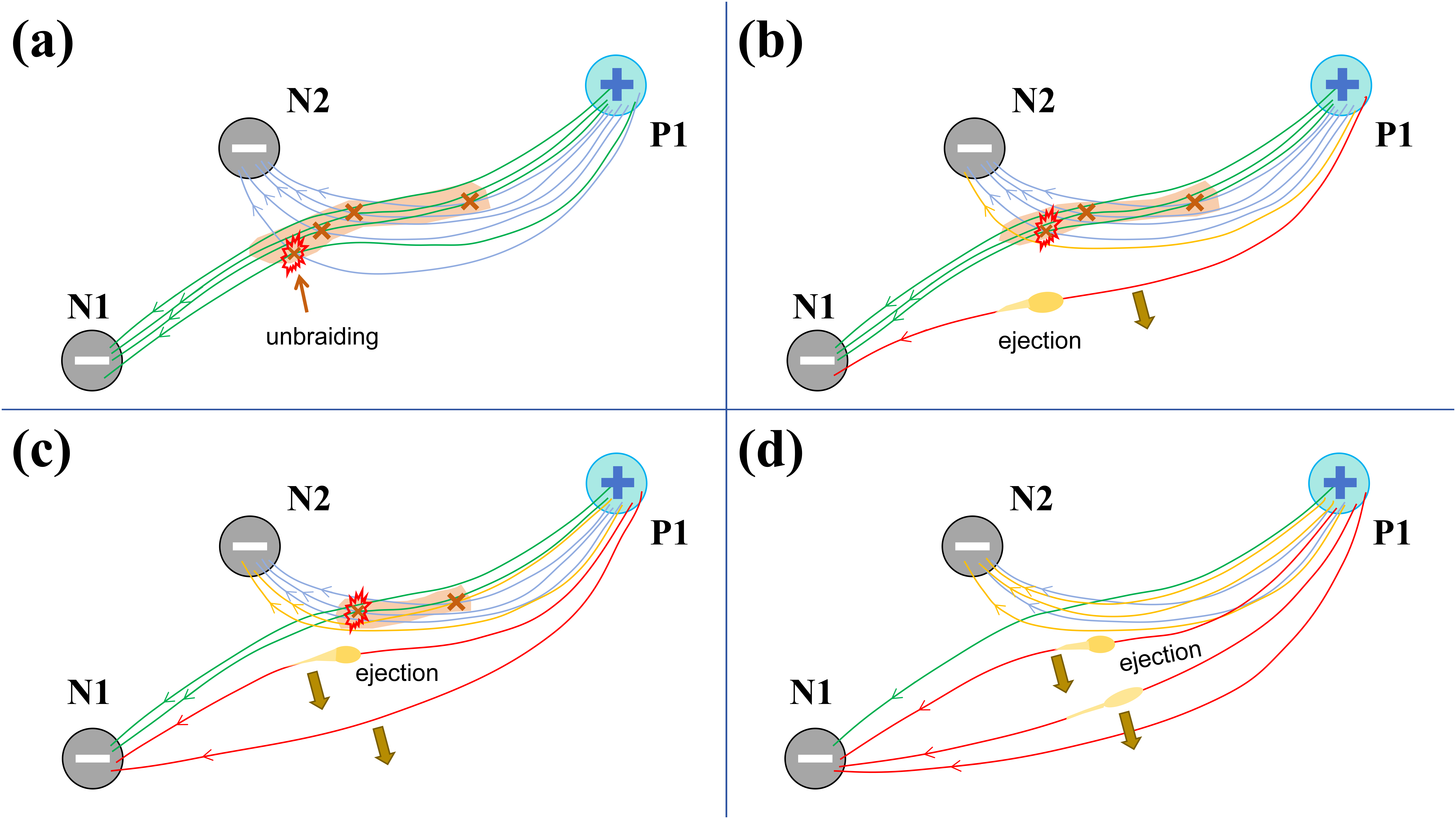}
		\caption{A schematic diagram to interpret the activities observed in the present study.
		The series of ejections of plasma blobs is generated by component magnetic reconnection/unbraiding processes between the long loops (green lines connecting N1 and P1) and the short loops (cyan lines connecting N2 and P1).
		Post-reconnection loops shown in red carry the plasma blobs due to their larger curvatures than those in gold.
	}
	\label{fig7}
\end{figure*}

\section{Results} \label{sec:res}
Figure\,\ref{fig1} shows the overview of the dynamic loops, with panel (d) depicting the primary regions involved. 
The activities occur in the loop systems connecting two magnetic features with negative polarity (``N1'' \& ``N2'') and another feature with positive polarity (``P1''), which can be seen in Figure\,\ref{fig1}(d)--(f).
These include long loops from ``LF1'' to ``LF3'' connecting ``N1'' and ``P1'' with lengths more than 65\,Mm and shorter ones from ``LF2'' to ``LF3'' connecting ``N2'' and ``P1'' with lengths of about 26\,Mm.
The loop system consists of many loop threads, which can be well distinguished in the HRI images (Figure\,\ref{fig1}(d)).
The west parts of the long loops and the shorter ones coincide apparently (see the green dotted lines in Figure\,\ref{fig1}(e)), which appear in a ``Y'' shape.
The activities occur at the locations where the two loop system converge (see the bright dots in the region pointed by the green arrow in Figure\,\ref{fig1}(d)).
The converged part of these loops has a width of about 7 pixels (i.e. 800\,km).
It is worthy to mention that loops connecting the polarities of ``N2'' and ``P2'' also exist and are visible later in the observations, however, they are unlikely linking to the activities studied here.

\par
In Figure\,\ref{fig2} and the associated animation, we zoom in the region of activities in HRI/EUI 174\,\AA\ observations.
Around 09:18:20\,UT, we observe the short coronal loops move northward slightly, forming a parallel alignment with the long loops, accompanied by brightenings (bright nodes) in the converged location.
Between 09:19:08\,UT and 09:21:08\,UT, we can observe a series of bright blobs emerging from the middle of the loop system.
These bright blobs seem to be ejected from the converged part of the loops and propagating in the direction being perpendicular to the main axis of the loop system.
Some of them clearly show as elongated structures, and they expand along the loops that are expelled from the bundle of the pre-existed loops.
The running-different images shown in Figure\,\ref{fig2} can clearly show the propagating directions of these blobs, where the blobs are shown as a bright feature followed by a dark one.
Based on these running-different images, we determine the separation angles between the main axis and the expelled loops at the moment when they are detached from each other and exhibit brightenings.
To quantify this, we first identify the most prominent bright features along the main axis and those along the expelled loop, 
then we apply linear fittings to these feature points to obtain the orientations of their extensions (yellow dashed lines in the right column of Figure\, \ref{fig2}). 
The seperation angle is defined as the intersection angle of the two fitted lines, and it is treated as the seperation angle of the braided field lines of the corresponding event. 
Clearly, we also observe propagation of brightenings (nodes) along the main axis of the converged loops as indicate as blue dashed line in Figure \ref{fig2}(a).
These activities seem to be consequences of unbraiding of these tangled loops. 
The ejections of bright blobs are likely due to slingshot effect of the unbraided loops\,\citep{2021NatAs...5...54A} and the propagation of brightenings along the ``main axis'' could be a composite results of heating at and shifting of the unbraiding sites.


\par
The first ejection of plasma blobs starts at 09:19:08\,UT, which is seen in HRI/EUI for a duration of 60 seconds, and designated as ejection-1 (Figure\,\ref{fig2}(a)\&(f)).
Ejection-1 is not a compact blob, and it rather consists of a cluster of plasma with at least four blobs being ejected continuously from a ``node'' on the tangled loops.
During ejection-1, bright ``nodes'' (indicating of the sources of the plasma blobs) are intermittently distributed along the ``main axis'' for approximately 7-Mm-long and gradually propagate along that axis.
With the HRI data, the spatial scales of the entire ejection-1 is 1624\,km$\times$1983\,km, in which each of the blobs is about 800\,km$\times$800\,km.
The separation angle between the main axis and the expelled loop is approximate 39\degree.
A time-slice map along the propagation direction of ejection-1 (see the purple dashed line in Figure\,\ref{fig2}(a)) is shown in Figure\,\ref{fig3}(a), showing that ejection-1 has a speed of about 87\,\kms.
The complex dynamics of ejection-1 can be due to disturbances of a loop system with multiple-braids, in which an unbraiding event can trigger the other braids\,\citep{2020ApJ...899...19C}.
During this process, magnetic free energy is continually converted into internal and kinematic energy, which energize the plasma blobs and the bright nodes. 



\par
At 09:20:08 UT, the second ejection of blob (ejection-2) occurs (see Figure\,\ref{fig2}(b)\&(g)).
It appears in only one frame of the HRI observations, and 
its size is measured to be 793\,km $\times$ 491 \,km.
The separation angle between the main axis and the expelled loop is approximate 26\degree.
Just 6 seconds before the appearance of ejection-2 (one HRI frame before), we observe a brightening in the node of the converged loops, and if we assume this brightening is the source, the speed of ejection-2 is estimated to be about 126\,\kms\ (Figure\,\ref{fig3}(b)).
It is worthy to mention that its speed has large uncertainty since it is seen in only one frame of the observations.
Based on its locations and occurring time, it is likely a continuation of ejection-1. 


\par
At 09:20:14 UT,  the converged part of the loops brightens noticeably and quickly produces a bright ejection of blobs (ejection-3).
This ejection lasts for 42 seconds and includes a bright head and a fade ``tail'' (Figure\,\ref{fig2}(c)\&(h)), indicating that the plasma ejection has an extended structure along the expelled loops.
Its core has a size of 793\,km$\times$704\,km, and its overall length including the tail is 2412\,km.
This observational feature suggests that the plasma ejection is still constrained by the magnetic field lines. 
The separation angle between the main axis and the expelled loop is approximate 25\degree.
Correspondingly, a compact brightening also appears in the main axis where this ejection starts.
Interestingly, ejection-3 extends towards only the loose end of the loops (towards the left in Figure\,\ref{fig2}(c)) but not the converged one (towards right in Figure\,\ref{fig2}(c)).
Since the tangled loops likely contain magnetic tips that can host cold plasma, it might block the heat flows and thus not visible in the HRI 174\,\AA\ images.
The time-slice maps along its propagating direction is shown in Figure\,\ref{fig3}(c), which gives a speed of 87\,\kms\ for its propagation.


\par
Following to ejection-3, at 09:20:32 UT, the bright node along the main axis of the loop system moves for a short distance and a cluster of bright blobs (ejection-4) is released (Figure\,\ref{fig2}(d)).
Ejection-4 lasts for about 30 seconds.
It shows similar morphology as ejection-3, with its length extending along the loops is about 1843\,km and that along its propagating direction is about 491\,km.
The separation angle between the main axis and the expelled loop is approximate 29\degree.
Its propagating speed derived from the time-slice map is 79\,\kms\ (Figure\,\ref{fig3}(d)). 


\par
At 09:20:38 UT,  there was a brightening at the node of the coronal loop, and 6 seconds later, another ejection of blobs (ejection-5) takes place (Figure\,\ref{fig2}(e)\&(j)).
In contrast to ejection-3 and ejection-4, ejection-5 is more compact, with a spatial scale of 886\,km$\times$737\,km.
The separation angle between the main axis and the expelled loop is approximate 35\degree.
Its propagation speed derived from the time-slice map is 82\,\kms\ (Figure\,\ref{fig3}(e)).

\par
When the ejections take place, we also observe propagation of bright features along the main axis (see the animation associated with Figure\,\ref{fig2}).
In Figure\,\ref{fig3}(f), we show the time-slice map along the main axis (the blue dashed line in Figure\,\ref{fig2}(a)).
It is clear that a flow along the main axis with a speed of about 45\,\kms\ is present during and after the ejections.
It is also clear that an intense brightening is coincident with ejections 2--5.
We would like to note that this flow about 30\,s after ejection-3 is mixed by another flow propagating from the east end of the loop system (``LF2'' in Figure\,\ref{fig1}(d) ).
Therefore whether the brightenings due to the ejections can propagate further toward the end of ``LF3'' is not clear.


\par
In AIA data, these ejections of plasma blobs are not visible, which is very likely due to the lower resolutions of the data and the viewing perspective of the instrument.
In contrast, the main loop-system including its converged part (bright node) are visible in the multi-wavelength data of AIA (see Figure\,\ref{fig1}(e)).
In Figure\,\ref{fig4} and the associated animation, we show the evolution of the region viewed in six AIA-EUV-channels.
The main loop-system can be clearly seen in the 171\,\AA, 193\,\AA, 211\,\AA\ and  304\,\AA\ passbands (Figure\,\ref{fig4}(b)--(e)), while the bright node also shows signatures in the 131\,\AA\ and 335\,\AA\ passbands (see the features pointed by the arrows in Figure\,\ref{fig4}).
In the animation, the dynamics in the bright node can also be seen, which is corresponding to those observed in the HRI/EUI observations.
In Figure\,\ref{fig5}, we show the AIA lightcurves of the bright node of the converged loops.
During the plasma ejections, the emission of the converged loops in these passbands is actually in a relatively low level.
There are some variations in these lightcurves and some local peaks are coincident with the plasma ejections, but we cannot confirm their one-to-one connections due to the low resolution of the AIA.
After the ejections, the emission of the converged loops are indeed enhanced in all the AIA EUV passbands and HRI/EUI 174\,\AA.
As described earlier, this enhancement is contributed by both these ejections and the other bright feature propagating from the east footpoint of ``LF2''.

\par
With multi-wavelength observations from AIA, we perform thermal analyses to the region using a Differential Emission Measure (DEM) algorithm\,\citep{2015ApJ...807..143C, 2018ApJ...856L..17S}, and the results are shown in Figure\,\ref{fig6} and the associated animation.
The maps shown in Figure\,\ref{fig6} are obtained from the observations at 09:25:46\,UT (i.e. 09:20:08\,UT for EUI observations when the ejections are taking place), in the middle when the ejections are taking place.
We can see that the node of the converged loops is visible in logT/K=5.5--5.8 and logT/K=6.1--6.4.
Based on the fact that the bright node can be clearly seen in the AIA 304\,\AA\ passband and also all the EUV passbands except AIA 94\,\AA, 
its temperature is most likely logT/K=5.5--5.8, 
because all the AIA EUV passbands have significant contributions from emission formed at logT/K=5.0--5.8\,\citep{2011A&A...535A..46D}.
From the animation, we see that the bright node shows some enhancement in the emission of logT/K=6.1--6.4, which suggests that it might actually be heated to these temperatures during the occurrence of the ejections.
Therefore, the main structure of the loops system should stay in temperatures of the upper transition region or lower corona.

\section{Discussion}
\label{sec:dis}
The activities observed here might be interpreted by the schematic diagrams illustrated in Figure\,\ref{fig7}.
In a loop system with tangled loops, component reconnection occur along the main axis,  and that results in heating in the loop system and also drives the ejections of plasma blob due to the slingshot effect of post-reconnection field lines\,\citep{2021NatAs...5...54A}.
Because the loop system consists of much more than two tangled loops, 
such activities can occur continuously due to persistent disturbances from the outside of the loop system or a domino effect, 
and thus produces the series of ejections presented in these observations.
We also notice that the typical bi-directional ejections corresponding to a magnetic reconnection geometry are not seen in the present cases. 
This is likely due to the geometry of the magnetic braids (Figure\,\ref{fig7}), 
where the post-reconnection field lines hosting the visible ejections have greater tensions than those extending along the main axis. 
This is similar to the cases reported by \citet{2021NatAs...5...54A}, who found the magnetic tension at the unbraiding site is highly asymmetric, with the ratio of inner to outer magnetic tension reaching as high as 100:1 and thus the scale of ejections at opposite directions are not equal. 
Additionally, the post-reconnection loops along the main axis are likely still braiding with the other loops and thus the flow is constrained near their original locations.

\par
Our observations demonstrate that such plasma ejections are constrained in loops rather than producing collimating jets.
This is significantly different from the nano-jets reported by \citet{2020ApJ...899...19C} and \citet{2021NatAs...5...54A}.
In the highly ionized plasma (i.e. the upper transition region and corona), we believe that ejected plasma due to a component reconnection of a magnetic braid should still be confined by magnetic field lines (i.e. loops) as shown in the present observations. 
The ``nano-jet'' phenomena reported by the two previous studies might also be expelled prominent loops but without clear extensions, because the cold and dense prominence-plasma are not easy to be heated and thus less pressure gradient along the loop leads to a small extension. 
Alternatively, the neutral particles in the prominence might allow cross-field motions and produce such a small and localized ejection.
Therefore, those ``nano-jets'' might not be jets of the classical definition. 
The difference between magnetic braiding in partialy ionized plasma and that in fully ionized one could be an important question for an in-depth investigation in the future.

\par
Assumed that these ejections of blobs are transition region structures and have typical upper-transition-region density ($n$) of $\sim1\times10^{10}$\,cm$^{-3}$,
for an ejected blob with a volume ($V$) of (700\,km)$^3$ and a speed ($v$) of 90\,\kms, which are typical values of the ejections observed here, the kinematic energy ($E_k$) expresses
$$E_k=\frac{1}{2}n m_p V v^2,$$
where $m_p$ is the mass of proton, and is obtained as $\sim2.3\times10^{23}$\,erg.
The geometric parameters of the loops here are similar to those listed as 4, 24 and 44 in Table 3 of \citet{2017ApJ...842...38X}, where they have magnetic strengths of 7\,G, 12\,G and 14\,G at their apexes, respectively.
Then, we estimate the magnetic field strength at the loop apex of the present case to be 10 G, though this remains a rough approximation. 
Therefore, the magnetic free energy ($E_m$) can be achieved by
$$E_m=\frac{(B\sin{\theta})^2}{8\pi}V,$$
where $B$ is the magnetic strength at the loop top, $\theta$ is the separation angle between the main axis and the expelled loops, and $V$ is the volume of the blob.
We can take a typical value of $\theta$=30\degree, and then $E_m\approx3.4\times10^{23}$\,erg.
This exactly falls in the regime of nano-flares.
The result means that a majority of the magnetic free energy goes into the kinematic energy rather than heating/radiation.
This also explains why the brightenings of the ejections are limited in a certain length along the loops.
The dissipation of the kinematic energy via other mechanisms, such as Kelvin-Helmholtz instabilities and K\'arm\'an vortex streets\,\citep{2023A&A...678L...7W}, should be taken into account seriously in the nanoflare heating mechanism. 
Furthermore, to understand the significance of such phenomena in the great coronal heating problem, a statistical study based on many more EUI observations targeting on regions with various activities in combination with high resolution magnetograms taken by PHI/SO should also be carried out in the future.

\section{Conclusions} \label{sec:con}
In the present work, we report on observations of activities associated with unbraiding of tangled loops that undergoes component magnetic processes in the solar atmosphere.
The unprecedentedly-high resolution observations of HRI/EUI reveal a series of ejections of plasma blobs produced from the converged part of the loop system.
Some of these plasma blobs show extensions along the loops expelled from the original loop system, indicating that they are constrained by the magnetic loops.
The bright cores of these ejection have a size of about 700\,km and durations less than 1 minute, and they travel in a speed of about 90\,\kms.
Their extensions along the post-reconnection loops are about 2\,000\,km.
The heat flows propagating along the main axis of the tangled loops have a speed of about 45\,\kms.
The separation angles between the post-reconnection loops and the main axis of the tangled loops are about 30\degree.
Thermal analyses based on multi-wavelength observations from AIA suggest the reconnection site most likely has temperatures in logT/K=5.5--5.8 and might heat to logT/K=6.1--6.4 occasionally.
The kinematic energy of such an ejection of blob is about $2.3\times10^{23}$\,erg, if we assume it has a density of the typical value of the upper transition region.

\par
We suggest the series of ejections of plasma blobs are resulted from the slingshot effect of post-reconnection loops in component magnetic reconnection.
Such a plasma ejection is more likely constrained in loops but not collimating jets.
The magnetic free energy in such a component magnetic reconnection is estimated to be $3.4\times10^{23}$\,erg, being consistent with the nanoflare heating scenario.
This indicates that a majority of the magnetic free energy in such a component reconnection is transferred into kinematic energy rather than heating or radiation.
The other dissipation mechanisms, such as instabilities that dissipate kinematic energy in the solar corona, are crucial in the nanoflare mechanism of coronal heating.
Our study not only provides another evidence for energy release by magnetic braids in the solar atmosphere, but also reveal new observational characteristics of such a process that are constraints for further theoretical studies.

\begin{acknowledgments}
We are grateful to the anonymous referee for the critical and constructive comments that help improve the manuscript.
This work is supported by the the National Key R\&D Program of China (2021YFA0718600) and National Natural Science Foundation of China (Nos. 42230203, 42174201, 42474223).
Project Supported by the Specialized Research Fund for State Key Laboratory of Solar Activity and Space Weather, Chinese Academy of Sciences.
Z.H. thanks ISSI-bern for supporting the team of ``Novel Insights Into Bursts; Bombs; and Brightenings in the Solar Atmosphere from Solar Orbiter''.
Solar Orbiter is a space mission of international collaboration between ESA and NASA, operated by ESA. The EUI instrument was built by CSL, IAS, MPS, MSSL/UCL, PMOD/WRC, ROB, LCF/IO with funding from the Belgian Federal Science Policy Office (BELSPO/PRODEX PEA 4000134088); the Centre National d'\'Etudes Spatiales (CNES); the UK Space Agency (UKSA); the Bundesministerium f\"ur Wirtschaft und Energie (BMWi) through the Deutsches Zentrum f\"ur Luft- und Raumfahrt (DLR); and the Swiss Space Office (SSO).
The AIA and HMI data are used by courtesy of NASA/SDO, the AIA and HMI teams and JSOC.
\end{acknowledgments}

\bibliography{references1}{}
\bibliographystyle{aasjournal}
\end{document}